\documentclass{article}

\usepackage{arxiv}

\usepackage[utf8]{inputenc} 
\usepackage[T1]{fontenc}    
\usepackage{hyperref}       
\usepackage{url}            
\usepackage{booktabs}       
\usepackage{amsfonts}       
\usepackage{nicefrac}       
\usepackage{microtype}      
\usepackage{lipsum}
\usepackage{graphicx}
\usepackage{natbib}

\begin{document}
\vspace*{0.35in}

\begin{flushleft}
{\Large
\textbf\newline{A spatio-temporal model to understand forest fires causality in Europe}
}
\newline
\\
Oscar Rodriguez de Rivera\textsuperscript{1,*},
Antonio López-Quílez\textsuperscript{2},
Marta Blangiardo\textsuperscript{3},
Martyna Wasilewska\textsuperscript{1}
\\
\bigskip
\bf{1}  Statistical Ecology @ Kent, National Centre for Statistical Ecology. School of Mathematics, Statistics and Actuarial Science, University of Kent. Canterbury, UK. 
\\
\bf{2}  Dept. Estadística i Investigació Operativa; Universitat de València. Valencia, Spain.
\\
\bf{2}  Faculty of Medicine, School of Public Health; Imperial College of London. London, UK.
\\
\bigskip
* o.ortega@kent.ac.uk

\end{flushleft}

\begin{abstract}
Forest fires are the outcome of a complex interaction between environmental factors, topography and socioeconomic factors \citep{bedi2012}. Therefore, understand causality and early prediction are crucial elements for controlling such phenomenon and saving lives.\par
The aim of this study is to build spatio-temporal model to understand causality of forest fires in Europe, at NUTS2 level between 2012 and 2016, using environmental and socioeconomic variables. We have considered a disease mapping approach, commonly used in small area studies to assess the spatial pattern and to identify areas characterised by unusually high or low relative risk.

\end{abstract}

\keywords{Hierarchical Bayesian models\and disease mapping\and integrated nested laplace approximation\and forest fires\and causality\and spatio-temporal model}

\section{Introduction}

Nowadays, wildfires have become one of the most significant disturbances worldwide \citep{flan2009, pech2010, paus2012, card2014, boer2017, moli2018}. The combination of a longer drought period and a higher woody biomass and flammability of dominant species creates an environment conducive to fire spread \citep{pino1998, mill2005}. Furthermore, vegetation pattern changes with the abandonment of traditional rural activities plays a direct role in the increase of fire severity and ecological and economic fire impacts \citep{flan2009, chuv2014}. Fire behavior exceeds most frequently firefighting capabilities and fire agencies have trouble in suppressing flames while providing safety for both firefighters and citizens \citep{wert2016}.\par
Many areas across the world have seen a rise in extreme fires in recent years. Those include South America and southern and western Europe. They also include unexpected places above the Arctic Circle, like the fires in Sweden during the summer of 2018 \citep{degr2013, euro2019}.\par
Extreme fire events, which are also referred to as “megafires”, are becoming frequent on a global scale; recent fires in Portugal, Greece, Amazone and other areas confirm this fact. There is not complete agreement on the term “megafires”, which often refers to catastrophic fire events in terms of human casualties, economic losses or both \citep{sanm2013b}.\par
Climate change will reduce fuel moisture levels from present values around the Mediterranean region and the region will become drier, increasing the weather-driven danger of forest fires. The countries in highest danger are Spain, Portugal, Turkey, Greece, parts of central and southern Italy, Mediterranean France and the coastal region of the Balkans, according to recent research of the Joint Research Centre (JRC) \citep{deri2017}.\par
Most international reports on biomass burning recognize the importance of the human factors in fire occurrence \citep{FAO2007}. Although fire is a natural factor in many ecosystems, human activities play a critical role in altering natural fire conditions, either by increasing ignitions \citep{leon2003}, or by suppressing natural fires \citep{john2001, keel1999}. Both factors are contradictory, and act mainly through the mixture of fire policy practices, on one hand, and land uses and demographic changes on the other. Most developed countries have maintained for several decades a fire suppression policy, which has lead to almost total fire exclusion. The long term impact of that policy has implied an alteration of traditional fire regimes, commonly by increasing average burn severity and size, as a result of higher fuel accumulation \citep{pyne2001}, although other authors are more critical about the real implication of fire suppression policy \citep{john2001}, or they tend to put more emphasis on the impact of climate changes \citep{west2006}. For developing countries, fire is still the most common tool for land clearing, and therefore it is strongly associated to deforestation, especially in Tropical areas \citep{coch1999, defr2002}. The traditional use of fire in shifting cultivation has turned in the last decades to permanent land use change, in favour of cropland and grasslands. In addition, fire is a traditional tool to manage permanent grasslands, which are burned annually to favour new shoots and improve palatability \citep{hobb1991, chuv2010}.\par
Global and local implications of changing natural fire circumstances have been widely recognized, with major effects on air quality, greenhouse gas emissions, soil degradation and vegetation succession \citep{goet2006, pari2006, rand2005}. The role of human activities in changing those conditions has not been assessed at global scale. Several local studies have identified factors that are commonly associated to human fire ignition, such as distance to roads, forest-agricultural or forest-urban interfaces, land use management, and social conflicts (unemployment, rural poverty, hunting disputes,) \citep{leon2003, mart2009, vega1995}. On the other hand, humans not only cause fires, but they suffer their consequences as well. Fire is recognized as a major natural hazard \citep{FAO2007}, which imply severe losses of human lives, properties and other socio-economic values \citep{rade2005, reis2006}.\par
Fire is no longer a significant part of the traditional systems of life; however, it remains strongly tied to human activity \citep{leon2009}. Knowledge of the causes of forest fires and the main driving factors of ignition is an indispensable step toward effective fire prevention \citep{gant2013}. It is widely recognized that current fire regimes are changing as a result of environmental and climatic changes (Pausas and Keeley 2009) with increased fire frequency in several areas in the Mediterranean Region of Europe \citep{rodr2013}. In Mediterranean-type ecosystems, several studies have indicated that these changes are mainly driven by fire suppression policies \citep{minn1983}, climate \citep{paus2012}, and human activities \citep{balm2011}. Human drivers mostly have a temporal dimension, which is why an historical/temporal perspective is often required \citep{zumb2011, carm2012}. In Mediterranean Europe, increases in the number of fires have been detected in some countries, including Portugal and Spain \citep{sanm2013a, rodr2013}. In addition, a recent work by \cite{turc2016} suggests huge spatial and temporal variability in fire frequency trends especially in the case of Spain, where increasing and decreasing trends were detected depending on the analysis period and scale. This increase in wildfire frequency and variability, with its associated risks to the environment and society \citep{moren2011, moren2014}, calls for a better understanding of the processes that control wildfire activity \citep{mass2013}. In recent decades, major efforts have been made to determine the influence of climate change on natural hazards, and to develop models and tools to properly characterize and quantify changes in climatic patterns. While physical processes involved in ignition and combustion are theoretically simple, understanding the relative influence of human factors in determining wildfire is an ongoing task \citep{mann2016}. It is clear that human-caused fires that occur repeatedly in a given geographical area are not simply reducible to individual personal factors, and thus subject to pure chance. They are usually the result of a spatial pattern, whose origin is in the interaction of environmental and socioeconomic conditions \citep{kout2015}. This is particularly true in human-dominated landscapes such as Spain, where anthropogenic ignitions surpass natural ignitions, and humans interact to a large degree with the landscape, changing its flammability, and act as fire initiators or suppressors. In such cases, human influence may cause sudden changes in fire frequency, intensity, and burned area size \citep{pezz2013}.\par
Fire is an integral component of Mediterranean ecosystems since at least the Miocene \citep{duba1995}. Although humans have used fires in the region for tens of thousands of years \citep{gore2004}, it is only in the last 10,000 or so that man has significantly influenced fire regime \citep{dani2010}. The use of fire as a management tool has persisted until these days, although the second half of the past century saw a major change and a regime shift due to abandonment of many unproductive lands \citep{moren1998, paus2012}. Although fire still is a traditional management tool in some rural areas for control of vegetation and enhancement of pastures for cattle feed, most fires these days are no longer related to the management of the land \citep{sanm2012, sanm2013a, sanm2013b}.\par
The European Mediterranean region is a highly populated area where nearly 200 Million people live in just 5 European Union countries, Portugal, Spain, France, Italy and Greece. Population density varies but remains very high with about 2500 inhabitants/km2 in the French Riviera (with peaks of up to 750,000 tourists per day during the summer)  \citep{cort2007} versus an average of 111 inhabitants/km2 in the region. The region is characterized by an extensive wildland urban interface (WUI). Large urban areas have expanded into the neighboring wildland areas, where expensive households are built. The WUI has been further increased by the construction of second holiday homes in the natural environment. Fire prone areas along the Mediterranean coast have been extensively built up, reducing in some cases the availability of fuels, but increasing largely the probability of fire ignition by human causes  \citep{gant2013}. In other areas of the same region, abandonment of the rural environment has lead to low utilization of forests, which are generally of limited productivity, and the subsequent accumulation of fuel loads  \citep{sanm2012, sanm2013a, sanm2013b, more2011}. The combination of the above factors converts the European Mediterranean region in a high fire risk area \citep{seba2008}, especially during the summer months when low precipitations and very high temperatures favor fire ignition and spread.\par
About 65,000 fires take place every year in the European region, burning, on average, around half a million ha of forest areas \citep{euro2011}. Approximately 85\% of the total burnt area occurs in the EU Mediterranean region \citep{sanm2010}. Although fires ignite and spread under favorable conditions of fuel availability and low moisture conditions, ignition is generally caused by human activities. Over 95\% of the fires in Europe are due to human causes. An analysis of fire causes show that the most common cause of fires is “agricultural practices”, followed by “negligence” and “arson” \citep{vila2009, reus2003}.\par
Most fires in the region are small, as a fire exclusion (extinction) policy prevails in Europe. Fires are thus extinguished as soon as possible, and only a small percentage escapes the initial fire attack and the subsequent firefighting operations. An enhanced international collaboration for firefighting exists among countries in the European Mediterranean region. This facilitates the provision of additional firefighting means to those in a given country from the neighbouring countries in case of large fire events. The trend of large fires, those larger than 500 ha, is shown quite stable in the last decades \citep{sanm2010}. However, among these large fires, several fire episodes caused catastrophic damages and the loss of human lives \citep{sanm2013a}.\par

A first step is to identify all the factors linked to human activity, establishing their relative importance in space and time  \citep{mart2009, mart2013}. According to  \cite{moren2014}, the number of fires over the past 50 years in Spain has increased, driven by climate and land-use changes. However, this tendency has been recently reversed due to fire prevention and suppression policies. This highlights the influence of changes in the role of human activities as some of the major driving forces. For instance, changes in population density patterns—both rural and urban—and traditional activities have been linked to an increase in intentional fires. In this sense, several works have previously investigated the influence of human driving factors of wildfires in Spain. These works have explored in detail a wide range of human variables  \citep{mart2009, chuv2010} and methods. Specifically, Generalized Linear Models  \citep{vila2009, mart2009, moren2014}, machine learning methods  \citep{lee1996, rori2014}, and more spatial-explicit models like Geographically Weighted Regression  \citep{mart2013, rodr2014} have previously been employed. However, all these approaches could be considered as stationary from a temporal point of view, since they are based on ‘static’ fire data information summarized or aggregated for a given time span. However, the influence of human drivers cannot be expected to be stationary  \citep{rodr2016}.  \cite{zumb2011} stress the importance of dealing with the temporal dimension of human drivers of wildfires. Therefore, exploring temporal changes in socioeconomic or anthropogenic drivers of wildfire will enhance our understanding of both current and future patterns of fire ignition, and thus help improve suppression and prevention policies  \citep{rodr2016}.\par
Disease risk mapping analyses can help to better understand the spatial variation of the disease, and allow the identification of important public health determinants  \citep{mora2018}. Spatio-temporal disease mapping models are a popular tool to describe the pattern of disease counts and to identify regions with an unusual incidence levels, time trend or both  \citep{schr2011}. This class of models is usually formulated within a hierarchical Bayesian framework with latent Gaussian model \citep{besa1991}. Several proposals have been made including a parametric  \citep{bern1995} and nonparametric  \citep{knor2000, laga2003, schm2004} formulation of the time trend and the respective space-time interactions. \par
Areal disease data often arise when disease outcomes observed at point level locations are aggregated over subareas of study region due to constraints such as population confidentiality. Producing disease risk estimates at area level is complicated by the fact that raw rates can be very unstable in areas with small population and for rare circumstances, an also by the presence of spatial autocorrelation that may exist due to spatially correlated risk factors  \citep{lero2000}. Thus, generalised linear mixed models are often used to obtain disease risk estimates since they enable to improve local estimates by accommodating spatial correlation and the effects of explanatory variables. Bayesian inference in these models can be performed using Integrated Nested Laplace Approximation (INLA) approach  \citep{rue2009} which is a computational alternative to the commonly used Markov chain Monte Carlo methods (MCMC); INLA allows to run fast approximate Bayesian inference in latent Gaussian models.\par
INLA is implemented in the INLA package for the R programming language, that provides an easy way to fit models via inla() function, which works in a similar way as other functions to fit models, such as glm() or gam()  \citep{palm2019}.\par
Statistical reporting in the European Union is done according to the Nomenclature of Units for Territorial Statistics (NUTS) system. The NUTS is a five-level hierarchical classification based on three regional levels and two local levels. Each member state is divided into a number of NUTS-1 regions, which in turn are divided into a number of NUTS-2 regions and so on. There are 78 NUTS-1 regions, 210 NUTS-2 and 1093 NUTS-3 units within the current 15 EU countries  \citep{euro2002}.\par
In this paper, we explore the application of these models to understand forest fires causality using environmental and socio-economic variables. We will work with areal data using the number of forest fires at NUTS-2 regional level in Europe and consider forest fires between 2012 and 2016.\par

\section{Material and Methods}

\begin{figure}[!ht]
    \centering
    \includegraphics[scale=0.6]{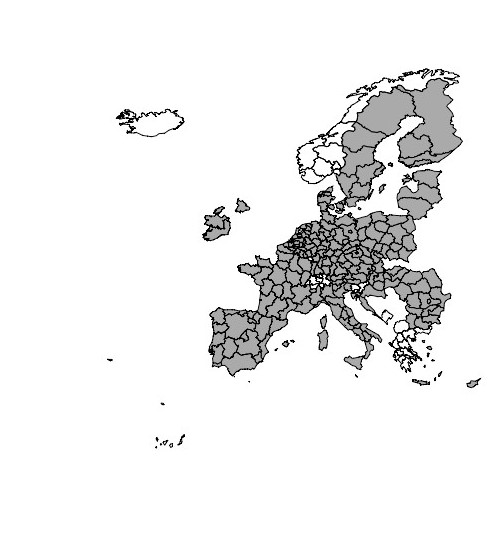}
    \caption{Study area, in grey administrative areas included in the analysis.}
    \label{fig1}
\end{figure}

We extend the analysis of globalization to the NUTS-2 regions of the 27 countries of the European Union (EU-27), as not all the regions have been included due to absence of information (forest fires or socio-economic data) (Figure \ref{fig1}).\par
Our main data set comprises the number of fires in Europe at NUTS-2 level, requested to the European Forest Fire Information System (EFFIS)  \citep{sanm2012}. We have chosen this level due to the variables that we are interested to analyse (socioeconomic and environmental).\par
\begin{figure}[!ht]
    \centering
    \includegraphics[scale=0.6]{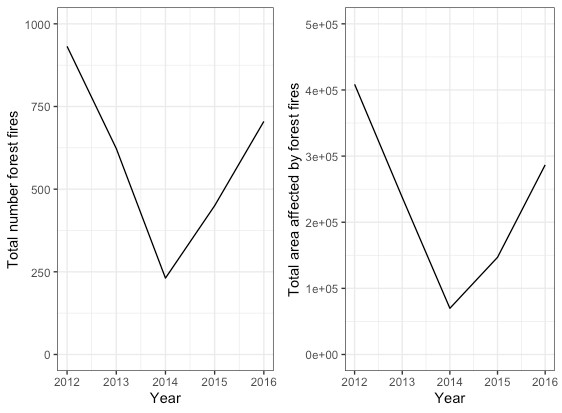}
    \caption{Summary of number of forest fires (left) and area affected between 2012 and 2016 (right)}
    \label{fig2}
\end{figure}

In order to summarise the forest fires in Europe we can see that the number of forest fires and the area affected have decreased between 2012 and 2014. However, the minimum was achieved in 2014, with a subsequent increase during 2015 and 2016 (Figure \ref{fig2}).\par
\begin{figure}[!ht]
    \centering
    \includegraphics[scale=0.4]{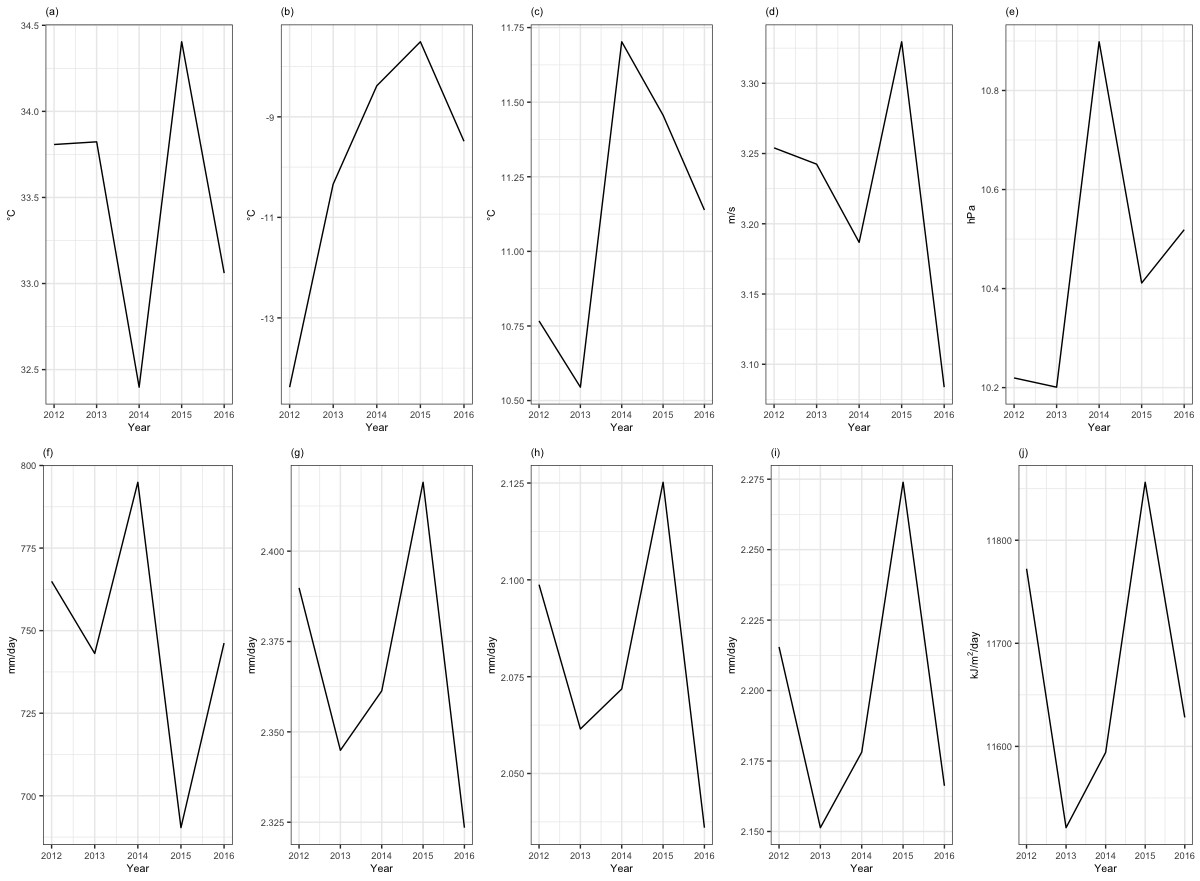}
    \caption{Trend of average of NUTS 2 regions by year of environmental variables between 2012 and 2015. (a) Maximum air temperature by year; (b) Miminum air temperature;(c) Mean air temperature; (d) Mean daily wind speed; (e) Vapour pressure; (f) Daily precipitation; (g) Potential evaporation from the water surface; (g) Potential evaporation from moist bare soil surface; (h) Potential evapotranspiration from crop canopy; (j)Total global radiation.}
    \label{fig3}
\end{figure}

The following environmental variables were obtained from the AGRI4CAST Resources Portal: Maximum air temperature ($^{\circ}$C); Minimum air temperature ($^{\circ}$C); Mean air temperature ($^{\circ}$C); Mean daily wind speed at 10 m. (m/s); Vapour pressure (hPa); Daily precipitation (mm/day); Potential evaporation from the water surface (mm/day); Potential evaporation from moist bare soil surface (mm/day); Potential evapotranspiration from crop canopy (mm/day); Total global radiation (kJ/m2/day). For each region we have the average by year. In Figure \ref{fig3} we can see the average of the different variables by year for all the NUTS 2 regions.\par
\begin{figure}[!ht]
    \centering
    \includegraphics[scale=0.4]{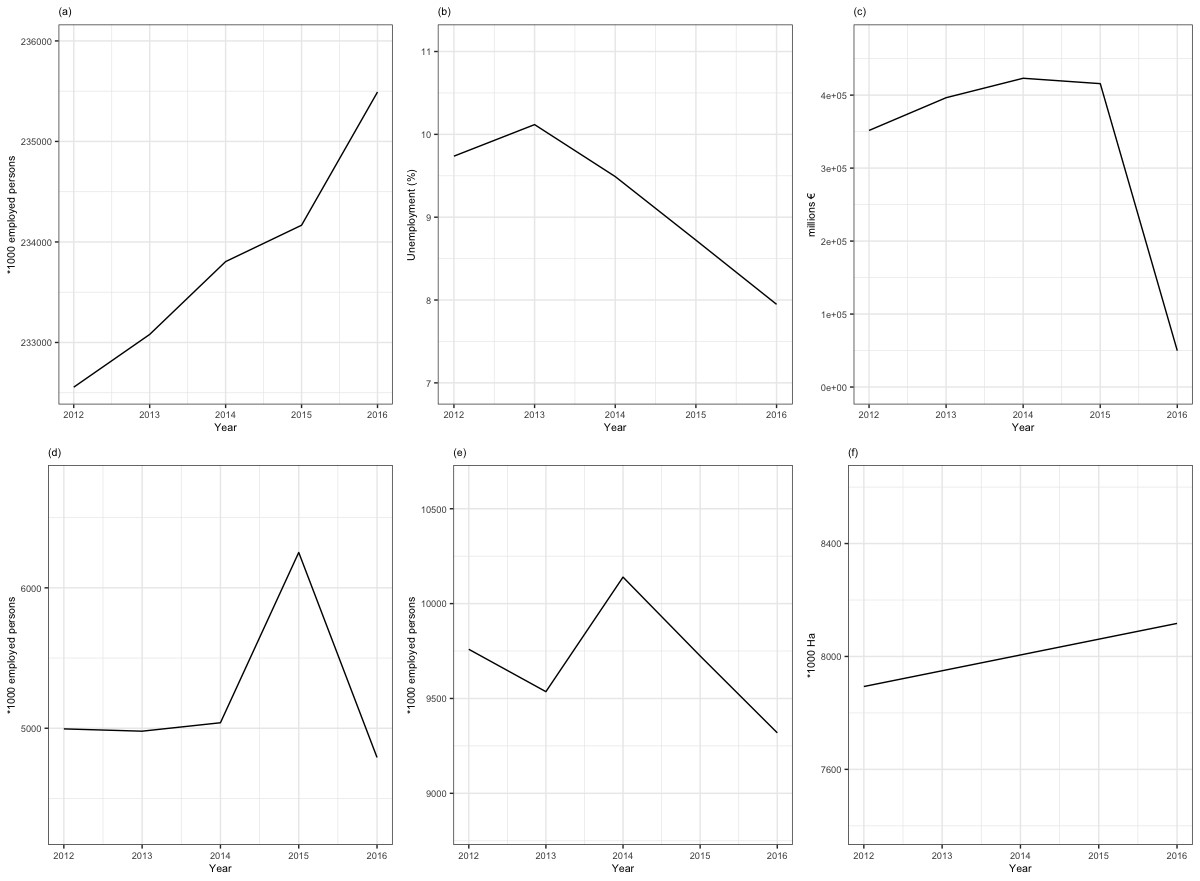}
    \caption{Trend the socioeconomic variables between 2012 and 2016. Totalisers by year in the following graphs: (a) Active population; (c) Economic aggregates of forestry; (d) Employed persons working in Forest sector; (e) Employed persons working in manufactured products from woodland); and (b) Average of Unemployment.}
    \label{fig4}
\end{figure}

The following socio-economic variables were obtained from Eurostat: Active population (*1000 employed persons), Woodland (*1000 hectares of Woodland in the area), Manufactured (*1000 employed persons working in manufactured products from woodland); Forestry (*1000 employed persons working in Forest sector); Economic aggregates of forestry (million euro) and Unemployment (\%). In this case we have included in our model totals values by year and region. In order to summarise the different variables we have included in Figure \ref{fig4} the total values by year for all the variables except for Unemployment where we have done the average for all regions by year.\par

\subsection{Spatio-Temporal model}

Here we consider a disease mapping approach, commonly used in small area studies to assess the spatial pattern of a particular outcome and to identify areas characterised by unusually high or low relative risk  \citep{laws2013, pasc2000}.\par
For the \emph{i-th} area, the number of forest fires $y_i$ is modelled as 

\begin{equation}
y_{it} \sim Poisson(\lambda_{it});  \lambda_{it} = E_{it} \rho_{it}
\end{equation}

where the $E_{it}$ are the expected number of forest fires and $\rho_{it}$ is the rate.\par
We specify a log-linear model on $\rho_{i}$ and include spatial, temporal and a space-time interaction, which would explain differences in the time trend for different areas. We use the following specification to explain these differences:

\begin{equation}
\rho_{it}=\alpha + \upsilon_{i}+ \nu_{i}+\gamma_{t}+\phi_{t}+\delta_{it},
\end{equation}

There are several ways to define the interaction term: here, we assume that the two unstructured effects $\nu_{i}$ and $\phi_{t}$ interact. We re-write the precision matrix as the product of the scalar $\tau_{\nu}$ (or $\tau_{\phi}$) and the so called structure matrix $\textbf{\emph{F}}_{\nu}$ (or  $\textbf{\emph{F}}_{\phi}$),  which identifies the neighboring structure; here the structure matrix $\textbf{\emph{F}}_{\delta}$ can be factorised as the Kronecker product of the structure matrix for $\nu$ and $\phi$ (Clayton, 1996): $\textbf{\emph{F}}_{\phi} = \textbf{\emph{F}}_{\nu} \otimes \textbf{\emph{F}}_{\phi}= \textbf{\emph{I}} \otimes \textbf{\emph{I}} = \textbf{\emph{I}}$ (because both $\nu$ and $\phi$ are unstructured). Consequently, we assume no spatial and/or temporal structure on the interaction and therefore $\delta_{it} \sim Normal(0,\tau_{\phi})$ — see Knorr-Held (2000) for a detailed description of other specifications.\par
In the model presented we assume the default specification of R-INLA for the distribution of the hyper-parameters; therefore, log$\tau_{\upsilon}$ $\sim$ logGamma(1,0.0005) and log$\tau_{\nu}$ $\sim$ logGamma(1,0.0005). In addition we specify a logGamma(1,0.0005) prior on the log-precision of the random walk and of the two unstructured effects \citep{blan2015}.\par
To evaluate the fit of this model, we have applied the Watanabe-Akaike information criterion (WAIC)  \citep{wata2010}. WAIC was suggested as an appropriate alternative for estimating the out-of-sample expectation in a fully Bayesian approach. This method starts with the computed log pointwise posterior predictive density and then adds a correction for the effective number of parameters to adjust for overfitting  \citep{gelm2013}. Watanabe-Akaike information criterion works on predictive probability density of detected variables rather than on model parameter; hence, it can be applied in singular statistical models (i.e. models with non-identifiable parameterization)  \citep{li2016}.\par
We have used Integrated Nested Laplace Approximation (INLA) implemented in R-INLA within the R statistical software (version 3.6.0).\par

\section{Results}

In this section, we show how the forest fires have evolved between 2012 and 2016.\par
\begin{figure}[!ht]
    \centering
    \includegraphics[scale=0.6]{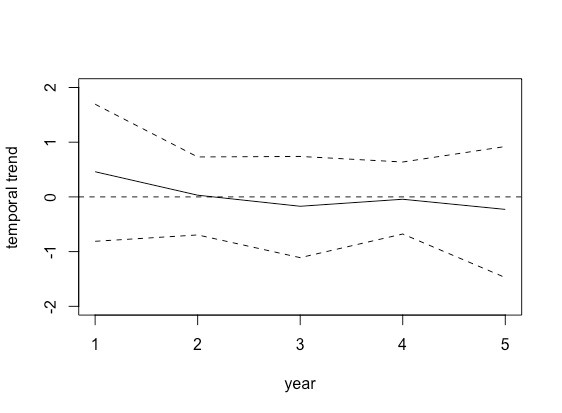}
    \caption{Global linear temporal trend for number of forest fires in Europe at NUTS2 region level. The solid line identifies the posterior mean for $\beta_{t}$ , while the dashed lines are the 95\% credibility intervals.}
    \label{fig5}
\end{figure}

Analysing the temporal trend, we can see graphically (Figure \ref{fig5}), the posterior temporal trend for forest fires in Europe. In this graph we show how the number of forest fires tend to be reduced over time.\par
\begin{figure}[!ht]
    \centering
    \includegraphics[scale=0.8]{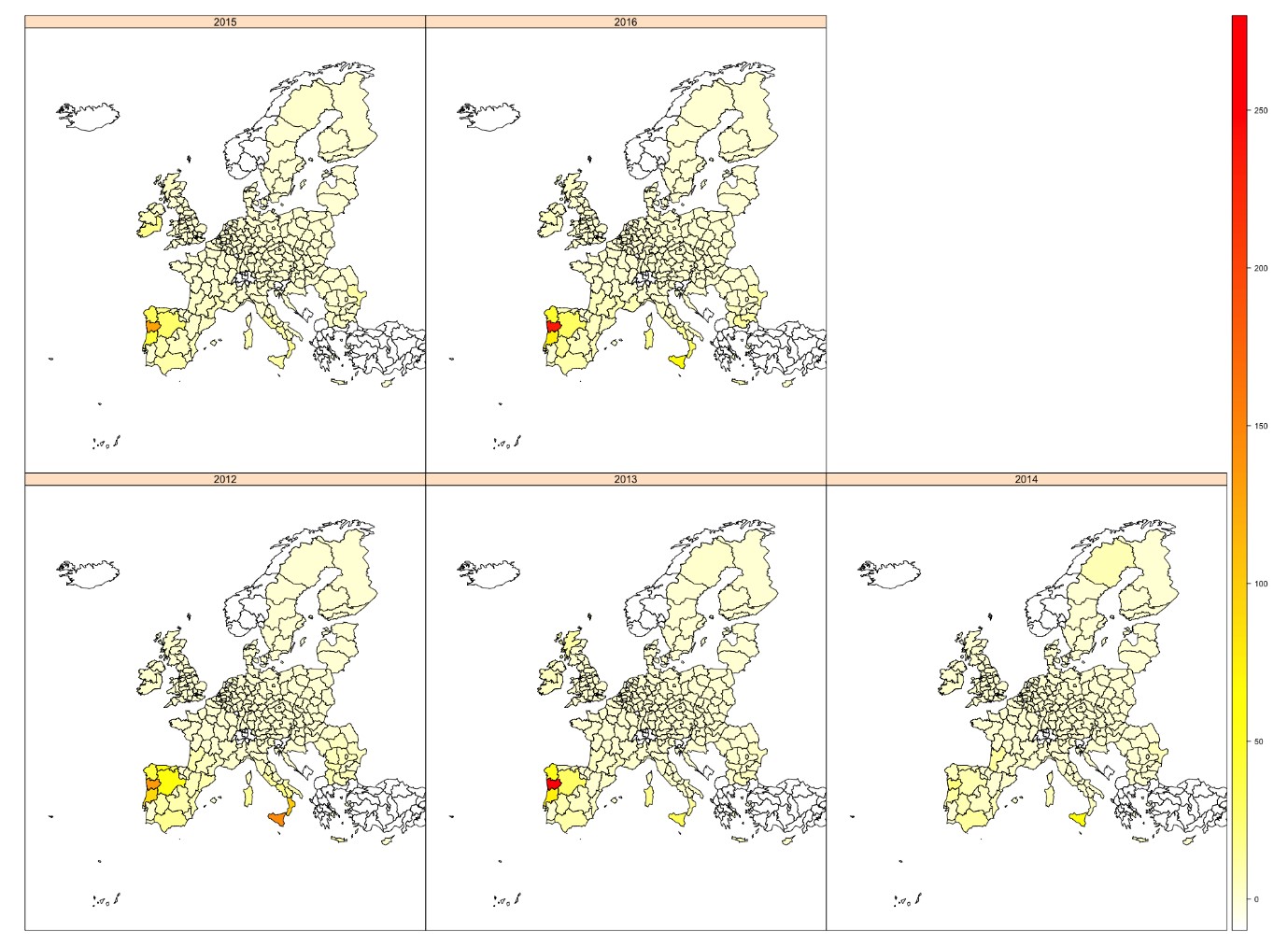}
    \caption{Map of the number of forest fires posterior distribution by region.}
    \label{fig6}
\end{figure}

Analysing the posterior distribution of forest fires (Figure \ref{fig6}) in Europe we can see that there is a “hot point” in western of the continent (North of Portugal and North West of Spanish peninsula). Also, as we can see, in general, the predicted number of forest fires is low in central Europe.\par
Comparing the different years, as we pointed previously, during 2014 the number of forest fires decreased in all areas except in some regions of Spain and Sicily. In addition, analysing the number of forest fires by region we can see that the region with stronger variations is the North region from Portugal.\par

\begin{figure}[!ht]
    \centering
    \includegraphics[scale=0.8]{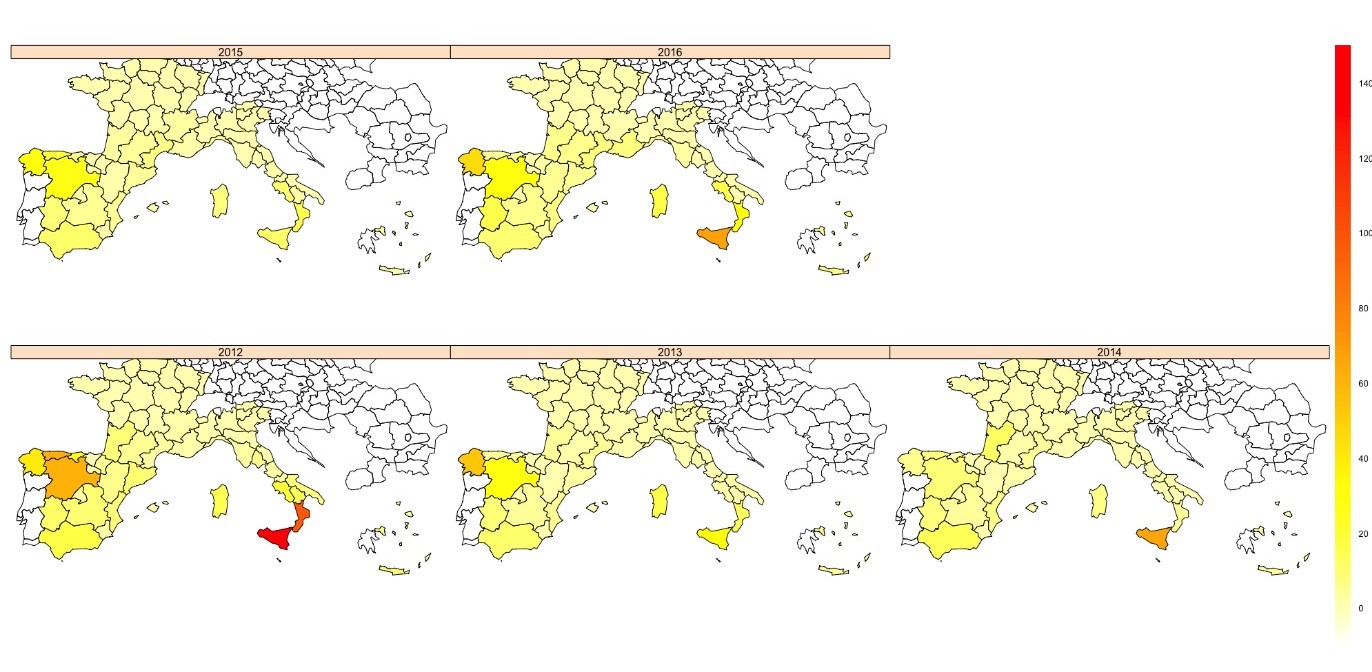}
    \caption{Detail of posterior distribution of forest fires in the Mediterranean region.}
    \label{fig7}
\end{figure}

In Figure \ref{fig7} we can see a more detailed map focused in Mediterranean countries (France, Greece, Italy and Spain). In this case it is clear that variability in France is almost inexistent only with some increase in the number of forest fires in Southern regions in 2016. The results from the data available for Greece, show that there are not big changes during the time analysed. However, Italy and Spain show more fluctuations during this period. The Southern part of Italy shows great changes along the time, starting with almost 150 forest fires in Sicily in 2012 to reduce until about 30 forest fires in 2015 and increase again in 2016 (67 forest fires). Similarly, in Spain the Northwest region shows several fluctuations. However, in Spain higher number of forest fires affects more regions. \par
As we can see in Table\ref{table1}, several variables are affecting the quantity of forest fires. But two of them have more impact (have higher values) than the others. Evaporation in water surface (EvaporationW) is affecting positively the volume of forest fires at region level. On the other hand, and with similar magnitude but negative sign, Evapotranspiration from crop canopy (Evapotrans.) is affecting negatively the presences of forest fires. \par

\begin{table}
 \caption{Posterior estimates summary (Mean, Standard deviation and 95\% Credible Interval). Fixed effects and hyperparameters for spatio-temporal model.}
  \centering
  \begin{tabular}{lllll}
    \toprule
    \multicolumn{1}{l}{Fixed effects}                   \\
    \cmidrule(r){1-5}
    	&mean	&sd	&0.025quant	&0.975quant \\
    \midrule
    Active	&0.3045	&0.1739	&-0.0375	&0.6468     \\
    Aggregates	&-0.2919	&0.1568	&-0.6008	&0.0152 \\
    Forestry	&0.6073	&0.3321	&-0.058	&1.2472  \\
    Manufactured	&-0.9303	&0.3944	&-1.717	&-0.1669\\
    MaxTemperature	&0.1761	&0.14	&-0.0999	&0.4503\\
    MinTemperature	&0.584	&0.2153	&0.1631	&1.0083\\
    AvgTemperature	&-0.1396	&0.4931	&-1.1094	&0.8277\\
    Wind	&0.5609	&0.2512	&0.0655	&1.0522\\
    Presion	&-0.28	&0.2978	&-0.8653	&0.3046\\
    Precipitation	&-0.0224	&0.0984	&-0.2158	&0.1707\\
    Evapotrans	&-23.6247	&9.8466	&-43.0572	&-4.3758\\
    EvaporationW	&24.5911	&10.907	&3.2597	&46.0829\\
    EvaporationS	&0.9008	&0.6832	&-0.4398	&2.2435\\
    Radiation	&-0.2677	&0.9264	&-2.0907	&1.5486\\
    Woodland	&0.7649	&0.2249	&0.3331	&1.2172\\
    \toprule
    \multicolumn{1}{l}{Model hyperparameters}                   \\
    \cmidrule(r){1-5}
    	&mean	&sd	&0.025quant	&0.975quant \\
    	\midrule
    Precision for AREA\_ID	&2.02E-01	&3.85E-02	&0.1347	&2.85E-01\\
    Precision for Year	&1.14E-01	&6.61E-02	&0.0299	&2.80E-01\\
    Precision for AREA\_ID.YEAR	&1.59E+00	&2.66E-01	&1.1262	&2.17E+00\\

    \bottomrule
  \end{tabular}
  \label{table1}
\end{table}

However, evapotranspiration from crop canopy is having a negative effect in forest fires quantity. In this group we need to highlight variables having more impact (higher values) than the others. Evaporation in water surface (EvaporationW) is affecting positively the volume of forest fires at region level. On the other hand, and with similar magnitude but negative sign, Evapotranspiration from crop canopy (Evapotrans) is affecting negatively the presences of forest fires. The rest of the variables that are affecting positively the amount of forest fires are Minimum temperature at 10 m. (MinTemperature) and Mean daily wind speed at 10 m (WIND). Finally, Manufactured is affecting negatively the quantity of forest fires.\par

\begin{figure}[!ht]
    \centering
    \includegraphics[scale=0.5]{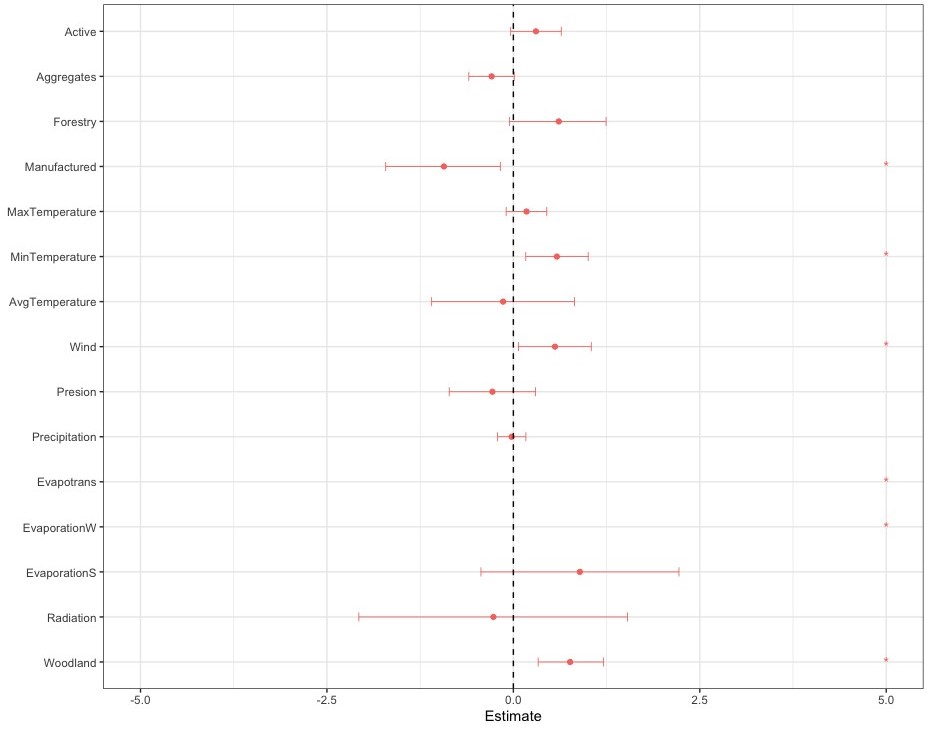}
    \caption{Graphical representation of fixed effecs. Evaporation in water surface (EvaporationW) and Evapotranspiration from crop canopy (Evapotrans) were excluded in order to obtain more detail of the rest of the variables.}
    \label{fig8}
\end{figure}

Graphical representation of estimation for the fixed effects is presented in Figure \ref{fig8}. This chart presents the variables and their relationships with forest fires. Variables distributed in a positive side contribute to higher number of forest fires; the opposite, with variables with negative distribution. Variables present in both areas (positive and negative) do not have a clear relationship with answer.\par

\section{Conclusions}

We have built spatio-temporal models to predict the quantity of forest fires in Europe at NUTS-2 regional level. We have shown the relationship between the different variables and the number of forest fires by region. We have shown that this relationship not only is between some of the variables (fixed effects), but also the evolution of forest fires along the time is affected not only by time and spatial effects but also by the combination of both (Precision for AREA\_ID.YEAR).\par
Initially our main objective in this project was to apply these models to Europe with more granularity, assuming that more local information will help to understand better the causality of forest fires. However due to data availability it was not possible to develop the project in that way. Currently not all the socioeconomic data is available for all the NUTS-3 regions in a continuous timestamp, being this characteristic necessary to carry a spatio-temporal analysis.\par
Also, several factors can affect in different ways depending of the area. In our case, variables have been assumed in a scale that in some of the cases local information can help to understand cause-effect of forest fires. \par
Analysing the models, we believe that the use of spatio-temporal models is an advantage for the understanding of the different dynamics, given that the temporal and spatio-temporal perspective is not very frequent analysing forest hazards. \par
Summarising, we can generalise that not only environmental factors but also socioeconomic variables are affecting the causality of forest fires. However, more data and more granularity in the analysis in needed in order to understand this causality.\par
Landscapes became more hazardous with the time, since land abandonment led to an increase in forest area. Treeless areas burned proportionally more than treed ones  \citep{urbi2019}. Fires in southern Europe have more preference shrublands than for forest types  \citep{more2011, oliv2014}, but may vary along locations  \citep{moren2011}. This could be due to a change in the ignition patterns owing to shifts in the wildland-agricultural and wildland-urban interfaces  \citep{rodr2014, modu2016}. The most vulnerable landscapes were those with diversity of land uses, with forest-agriculture mixtures  \citep{orte2012}. For these reasons, inclusion of vegetation to analyse causality needs to be studied.\par
Fire trends can be affected by changes in ignition cause. In European Mediterranean countries, a minor percentage of fires are caused by lightning, and most are caused by people. Fires of these two sources tend to occur at different locations  \citep{vazq1998}, which could affect the vegetation they burn and the difficulty of extinction. However, no changes between these two sources have been observed  \citep{gant2013}. Regarding people-caused fires, the majority of them are voluntary, followed by negligence \citep{urbi2019}. In recent times, negligence fires are increasing and voluntary ones decreasing \citep{gant2013}. Whether this is differentially affecting the number of fires trends, is something that needs research \citep{urbi2019}.\par
Spatio-temporal models and the R-INLA package appear to offer additional benefits beyond the traditional analysis used to understand the causes of this hazards. The combination of using a complex spatial latent field to capture spatial processes and an underlying simple additive regression model for the response variables relationship to the different factors, means that the fixed effects are potentially more straightforward to interpret \citep{gold2016}. R-INLA models are extremely flexible in their specifications, with spatial autocorrelation and observer bias being straightforwardly incorporated as random effects, while standard error distributions, such as Gaussian, Poisson, Binomial, and a variety of zero-inflated models, can be used interchangeably \citep{rue2009}. \par


\clearpage

\begin{thebibliography}{999}
\bibliographystyle{chicago}

\bibitem[Bal \emph{et al}(2011)]{balm2011}
Bal, M.C., Pelachs, A., Perez-Obiol, R., Julia, R. and Cunill, R., 2011. Fire history and human activities during the last 3300 cal yr BP in Spain's Central Pyrenees: the case of the Estany de Burg. Palaeogeography, Palaeoclimatology, Palaeoecology, 300(1-4), pp.179-190.

\bibitem[Biavetti \emph{et al}(2014)]{biav2014}
Biavetti I, Karetsos S, Ceglar A, Toreti A, Panagos P. 2014. European meteorological data: contribution to research, development, and policy support, Proc. SPIE 9229, Second International Conference on Remote Sensing and Geoinformation of the Environment (RSCy2014), 922907 (12 August 2014); https://doi.org/10.1117/12.2066286 \par

\bibitem[Bedia \emph{et al}(2014)]{bedi2012}
*Bedia, J., Herrera, S., Gutiérrez, J.M., Zavala, G., Urbieta, I.R. and Moreno, J.M., 2012. Sensitivity of fire weather index to different reanalysis products in the Iberian Peninsula. Natural Hazards and Earth System Sciences, 12(3), pp.699-708.\par

\bibitem[Bernardinelli \emph{et al}(2014)]{bern1995}
*Bernardinelli, L., Clayton, D., Pascutto, C., Montomoli, C., Ghislandi, M. and Songini, M., 1995. Bayesian analysis of space—time variation in disease risk. Statistics in medicine, 14(21‐22), pp.2433-2443.\par

\bibitem[Besag \emph{et al}(1991)]{besa1991}
Besag J, York J, Mollie A. 1991. Bayesian image restoration with two applications in spatial statistics. Annals of the Institute of Statistical Mathematics 43(1):1–59.\par

\bibitem[Blangiardo and Cameletti(2015)]{blan2015}
Blangiardo, M. and Cameletti, M., 2015. Spatial and spatio-temporal Bayesian models with R-INLA. John Wiley \& Sons.\par

\bibitem[Boer \emph{et al}(2017)]{boer2017}
Boer, M.M., Nolan, R.H., De Dios, V.R., Clarke, H., Price, O.F. and Bradstock, R.A., 2017. Changing weather extremes call for early warning of potential for catastrophic fire. Earth's Future, 5(12), pp.1196-1202.\par

\bibitem[Cardil \emph{et al}(2014)]{card2014}
Cardil, A., Molina, D.M. and Kobziar, L.N., 2014. Extreme temperature days and their potential impacts on southern Europe. Natural Hazards and Earth System Sciences, 14(11), pp.3005-3014.\par

\bibitem[Carmona \emph{et al}(2012)]{carm2012}
Carmona, A., González, M.E., Nahuelhual, L. and Silva, J., 2012. Spatio-temporal effects of human drivers on fire danger in Mediterranean Chile. Bosque, 33(3), pp.321-328.\par

\bibitem[Chuvieco \emph{et al}(2010)]{chuv2010}
Chuvieco, E., Aguado, I., Yebra, M., Nieto, H., Salas, J., Martín, M.P., Vilar, L., Martínez, J., Martín, S., Ibarra, P. and De La Riva, J., 2010. Development of a framework for fire risk assessment using remote sensing and geographic information system technologies. Ecological Modelling, 221(1), pp.46-58.\par

\bibitem[Chuvieco \emph{et al}(2014)]{chuv2014}
Chuvieco, E., Martínez, S., Román, M.V., Hantson, S. and Pettinari, M.L., 2014. Integration of ecological and socio‐economic factors to assess global vulnerability to wildfire. Global Ecology and Biogeography, 23(2), pp.245-258.\par

\bibitem[Cochrane \emph{et al}(1999)]{coch1999}
Cochrane, M.A., Alencar, A., Schulze, M.D., Souza, C.M., Nepstad, D.C., Lefebvre, P. and Davidson, E.A., 1999. Positive feedbacks in the fire dynamic of closed canopy tropical forests. Science, 284(5421), pp.1832-1835.\par

\bibitem[Corteau(2007)]{cort2007}
Corteau, R., 2007. Report No. 117 (2007–2008) for the French Parliament Office for the Evaluation of Scientific and Technological Choices, 60p.\par

\bibitem[Daniau \emph{et al}(2010)]{dani2010}
Daniau, A.L., d'Errico, F. and Goñi, M.F.S., 2010. Testing the hypothesis of fire use for ecosystem management by Neanderthal and Upper Palaeolithic modern human populations. Plos one, 5(2), p.e9157.\par

\bibitem[de Groot \emph{et al}(2013)]{degr2013}
de Groot, W.J., Flannigan, M.D. and Cantin, A.S., 2013. Climate change impacts on future boreal fire regimes. Forest Ecology and Management, 294, pp.35-44.\par

\bibitem[De Rigo \emph{et al}(2017)]{deri2017}
De Rigo, D., Libertà, G., Houston Durrant, T., Artés Vivancos, T. and San-Miguel-Ayanz, J., 2017. Forest fire danger extremes in Europe under climate change: variability and uncertainty. European Union: Luxembourg.\par

\bibitem[DeFries \emph{et al}(2002)]{defr2002}
DeFries, R.S., Houghton, R.A., Hansen, M.C., Field, C.B., Skole, D. and Townshend, J., 2002. Carbon emissions from tropical deforestation and regrowth based on satellite observations for the 1980s and 1990s. Proceedings of the National Academy of Sciences, 99(22), pp.14256-14261.\par

\bibitem[Dubar \emph{et al}(1995)]{duba1995}
Dubar, M., Ivaldi, J.P. and Thinon, M., 1995. Mio-pliocene fire sequences in the valensole basin (Southern France)-paleoclimatic and paleogeographic interpretation. Comptes Rendus De L Academie Des Sciences Serie Ii, 320(9), pp.873-879.\par

\bibitem[European Commission(2011)]{euro2011}
European Commission, 2011. Forest Fires in Europe 2010. Official Publication of the European Communities, EUR 24910.\par

\bibitem[European Commission(2019)]{euro2019}
EU parliament’s debate: Climate change and forest fires in Europe. Available online: https://eustafor.eu/climate-change-and-forest-fires-in-europe/ (accessed on 10 September 2019).\par

\bibitem[Eurostat(2002)]{euro2002}
European Commission. Eurostat database. 2019. http://ec.europa.eu/eurostat/Eurostat, 2002. Main characteristics of the NUTS. Available from: http://europa.eu.int/comm/eurostat/ramon/nuts/mainchar\_regions\_en.html.\par

\bibitem[Food(2007)]{FAO2007}
Food, U.N., 2007. Fire management–Global assessment 2006.\par

\bibitem[Flannigan \emph{et al}(2009)]{flan2009}
Flannigan, M.D., Krawchuk, M.A., de Groot, W.J., Wotton, B.M. and Gowman, L.M., 2009. Implications of
changing climate for global wildland fire. International journal of wildland fire, 18(5), pp.483-507.\par

\bibitem[Ganteaume \emph{et al}(2013)]{gant2013}
Ganteaume, A., Camia, A., Jappiot, M., San-Miguel-Ayanz, J., Long-Fournel, M. and Lampin, C., 2013. A review of the main driving factors of forest fire ignition over Europe. Environmental management, 51(3), pp.651-662.\par

\bibitem[Garcia \emph{et al}(1995)]{garc1995}
Garcia, C.V., Woodard, P.M., Titus, S.J., Adamowicz, W.L. and Lee, B.S., 1995. A logit model for predicting the daily occurrence of human caused forest-fires. International Journal of Wildland Fire, 5(2), pp.101-111.\par

\bibitem[Gelman and Shalizi(2013)]{gelm2013}
Gelman, A. and Shalizi, C.R., 2013. Philosophy and the practice of Bayesian statistics. British Journal of Mathematical and Statistical Psychology, 66(1), pp.8-38.\par

\bibitem[Goetz \emph{et al}(2006)]{goet2006}
Goetz, S.J., Fiske, G.J. and Bunn, A.G., 2006. Using satellite time-series data sets to analyze fire disturbance and forest recovery across Canada. Remote Sensing of Environment, 101(3), pp.352-365.\par

\bibitem[Golding and Purse(2016)]{gold2016}
Golding, N. and Purse, B.V., 2016. Fast and flexible Bayesian species distribution modelling using Gaussian processes. Methods in Ecology and Evolution, 7(5), pp.598-608.\par

\bibitem[Goren-Inbar \emph{et al}(2004)]{gore2004}
Goren-Inbar, N., Alperson, N., Kislev, M.E., Simchoni, O., Melamed, Y., Ben-Nun, A. and Werker, E., 2004. Evidence of hominin control of fire at Gesher Benot Yaaqov, Israel. Science, 304(5671), pp.725-727.\par

\bibitem[Hobbs \emph{et al}(1991)]{hobb1991}
Hobbs N.T., Schimel D.S., Owensby C.E., Ojima D.S., 1991. Fire and grazing in the tallgrass prairie – contingent effects on nitrogen budgets. Ecology 72:1374–1382. \par

\bibitem[Johnson \emph{et al}(2001)]{john2001}
Johnson, E. A., Miyanishi, K., \& Bridge, S. R. J., 2001. Wildfire regime in the boreal forest and the idea of suppression and fuel buildup. Conservation Biology, 15(6), 1554-1557.\par

\bibitem[Keeley \emph{et al}(1999)]{keel1999}
Keeley, J.E., Fotheringham, C.J. and Morais, M., 1999. Reexamining fire suppression impacts on brushland fire regimes. Science, 284(5421), pp.1829-1832.\par

\bibitem[Knorr‐Held(2000)]{knor2000}
Knorr‐Held, L., 2000. Bayesian modelling of inseparable space‐time variation in disease risk. Statistics in medicine, 19(17‐18), pp.2555-2567.\par

\bibitem[Koutsias \emph{et al}(2015)]{kout2015}
Koutsias, N., Allgöwer, B., Kalabokidis, K., Mallinis, G., Balatsos, P. and Goldammer, J.G., 2015. Fire occurrence zoning from local to global scale in the European Mediterranean basin: implications for multi-scale fire management and policy. iForest-Biogeosciences and Forestry, 9(2), p.195.\par

\bibitem[Lagazio \emph{et al}(2003)]{laga2003}
Lagazio, C., Biggeri, A. and Dreassi, E., 2003. Age–period–cohort models and disease mapping. Environmetrics: The official journal of the International Environmetrics Society, 14(5), pp.475-490.

\bibitem[Lawson(2013)]{laws2013}
Lawson, A.B., 2013. Bayesian disease mapping: hierarchical modeling in spatial epidemiology. Chapman and Hall/CRC.\par

\bibitem[Lee \emph{et al}(1996)]{lee1996}
Lee, B.S., Woodard, P.M. and Titus, S.J., 1996. Applying neural network technology to human-caused wildfire occurrence prediction. AI applications.\par

\bibitem[Leone \emph{et al}(2003)]{leon2003}
Leone, V., Koutsias, N., Martínez, J., Vega-García, C., Allgöwer, B. and Lovreglio, R., 2003. The human factor in fire danger assessment. In Wildland Fire Danger Estimation and Mapping: The Role of Remote Sensing Data (pp. 143-196).\par

\bibitem[Leone \emph{et al}(2009)]{leon2009}
Leone, V., Lovreglio, R., Martín, M.P., Martínez, J. and Vilar, L., 2009. Human factors of fire occurrence in the Mediterranean. In Earth observation of wildland fires in Mediterranean ecosystems (pp. 149-170). Springer, Berlin, Heidelberg.\par

\bibitem[Leroux \emph{et al}(2000)]{lero2000}
Leroux, B.G., Lei, X. and Breslow, N., 2000. Estimation of disease rates in small areas: a new mixed model for spatial dependence. In Statistical models in epidemiology, the environment, and clinical trials (pp. 179-191). Springer, New York, NY.\par

\bibitem[Li \emph{et al}(2016)]{li2016}
Li, L., Qiu, S., Zhang, B. and Feng, C.X., 2016. Approximating cross-validatory predictive evaluation in Bayesian latent variable models with integrated IS and WAIC. Statistics and Computing, 26(4), pp.881-897.\par

\bibitem[Mann \emph{et al}(2016)]{mann2016}
Mann, M.L., Batllori, E., Moritz, M.A., Waller, E.K., Berck, P., Flint, A.L., Flint, L.E. and Dolfi, E., 2016. Incorporating anthropogenic influences into fire probability models: Effects of human activity and climate change on fire activity in California. PLoS One, 11(4), p.e0153589.\par

\bibitem[Martínez \emph{et al}(2009)]{mart2009}
Martínez, J., Vega-Garcia, C. and Chuvieco, E., 2009. Human-caused wildfire risk rating for prevention planning in Spain. Journal of environmental management, 90(2), pp.1241-1252.\par

\bibitem[Martínez \emph{et al}(2013)]{mart2013}
Martínez-Fernández, J., Chuvieco, E. and Koutsias, N., 2013. Modelling long-term fire occurrence factors in Spain by accounting for local variations with geographically weighted regression. Natural Hazards and Earth System Sciences, 13(2), pp.311-327.\par

\bibitem[Massada \emph{et al}(2013)]{mass2013}
Massada, A.B., Syphard, A.D., Stewart, S.I. and Radeloff, V.C., 2013. Wildfire ignition-distribution modelling: a comparative study in the Huron–Manistee National Forest, Michigan, USA. International journal of wildland fire, 22(2), pp.174-183.\par

\bibitem[Millán \emph{et al}(2005)]{mill2005}
Millán, M.M., Estrela, M.J., Sanz, M.J., Mantilla, E., Martín, M., Pastor, F., Salvador, R., Vallejo, R., Alonso, L., Gangoiti, G. and Ilardia, J.L., 2005. Climatic feedbacks and desertification: the Mediterranean model. Journal of Climate, 18(5), pp.684-701.\par

\bibitem[Minnich(1983)]{minn1983}
Minnich, R.A., 1983. Fire mosaics in southern California and northern Baja California. Science, 219(4590), pp.1287-1294.\par

\bibitem[Modugno \emph{et al}(2016)]{modu2016}
Modugno, S., Balzter, H., Cole, B. and Borrelli, P., 2016. Mapping regional patterns of large forest fires in Wildland–Urban Interface areas in Europe. Journal of environmental management, 172, pp.112-126.\par

\bibitem[Molina \emph{et al}(2018)]{moli2018}
Molina, J.R., Moreno, R., Castillo, M. and y Silva, F.R., 2018. Economic susceptibility of fire-prone landscapes in natural protected areas of the southern Andean Range. Science of the Total Environment, 619, pp.1557-1565.\par

\bibitem[Moraga(2018)]{mora2018}
Moraga, P., 2018. Small Area Disease Risk Estimation and Visualization Using R. R J, 10, pp.495-506.\par

\bibitem[Moreira \emph{et al}(2011)]{more2011}
Moreira, F., Viedma, O., Arianoutsou, M., Curt, T., Koutsias, N., Rigolot, E., Barbati, A., Corona, P., Vaz, P., Xanthopoulos, G. and Mouillot, F., 2011. Landscape–wildfire interactions in southern Europe: implications for landscape management. Journal of environmental management, 92(10), pp.2389-2402.\par

\bibitem[Moreno \emph{et al}(2014)]{moren2014}
Moreno, M.V., Conedera, M., Chuvieco, E. and Pezzatti, G.B., 2014. Fire regime changes and major driving forces in Spain from 1968 to 2010. Environmental Science \& Policy, 37, pp.11-22.\par

\bibitem[Moreno \emph{et al}(2011)]{moren2011}
Moreno, J.M., Zuazua, E., Pérez, B., Luna, B., Velasco, A. and Resco de Dios, V., 2011. Rainfall patterns after fire differentially affect the recruitment of three Mediterranean shrubs. Biogeosciences, 8(12), pp.3721-3732.\par

\bibitem[Moreno \emph{et al}(1988)]{moren1998}
Moreno, J.M., Vázquez, A. and Vélez, R., 1998. Recent history of forest fires in Spain. Large forest fires, pp.159-185.\par

\bibitem[Oliveira \emph{et al}(2014)]{oliv2014}
Oliveira, S., Pereira, J.M., San-Miguel-Ayanz, J. and Lourenço, L., 2014. Exploring the spatial patterns of fire density in Southern Europe using Geographically Weighted Regression. Applied Geography, 51, pp.143-157.\par

\bibitem[Ortega \emph{et al}(2012)]{orte2012}
Ortega, M., Saura, S., González-Avila, S., Gómez-Sanz, V. and Elena-Rosselló, R., 2012. Landscape vulnerability to wildfires at the forest-agriculture interface: half-century patterns in Spain assessed through the SISPARES monitoring framework. Agroforestry systems, 85(3), pp.331-349.\par

\bibitem[Palmi-Perales \emph{et al}(2019)]{palm2019}
Palmi-Perales, F., Gomez-Rubio, V. and Martinez-Beneito, M.A., 2019. Bayesian Multivariate Spatial Models for Lattice Data with INLA. arXiv preprint arXiv:1909.10804.\par

\bibitem[Parisien \emph{et al}(2006)]{pari2006}
Parisien, M.A., Peters, V.S., Wang, Y., Little, J.M., Bosch, E.M. and Stocks, B.J., 2006. Spatial patterns of forest fires in Canada, 1980–1999. International Journal of Wildland Fire, 15(3), pp.361-374.\par

\bibitem[Pascutto \emph{et al}(2000)]{pasc2000}
Pascutto, C., Wakefield, J.C., Best, N.G., Richardson, S., Bernardinelli, L., Staines, A. and Elliott, P., 2000. Statistical issues in the analysis of disease mapping data. Statistics in medicine, 19(17‐18), pp.2493-2519.\par

\bibitem[Pausas \emph{et al}(2012)]{paus2012}
Pausas, J.G. and Fernández-Muñoz, S., 2012. Fire regime changes in the Western Mediterranean Basin: from fuel-limited to drought-driven fire regime. Climatic change, 110(1-2), pp.215-226.\par

\bibitem[Pausas and Keeley(2009)]{paus2009}
Pausas, J.G. and Keeley, J.E., 2009. A burning story: the role of fire in the history of life. Bioscience, 59(7), pp.593-601.\par

\bibitem[Pechony and Shindell(2010)]{pech2010}
Pechony, O. and Shindell, D.T., 2010. Driving forces of global wildfires over the past millennium and the forthcoming century. Proceedings of the National Academy of Sciences, 107(45), pp.19167-19170.\par

\bibitem[Pezzatti \emph{et al}(2013)]{pezz2013}
Pezzatti, G.B., Zumbrunnen, T., Bürgi, M., Ambrosetti, P. and Conedera, M., 2013. Fire regime shifts as a consequence of fire policy and socio-economic development: an analysis based on the change point approach. Forest Policy and Economics, 29, pp.7-18.\par

\bibitem[Piñol \emph{et al}(1998)]{pino1998}
Piñol, J., Terradas, J. and Lloret, F., 1998. Climate warming, wildfire hazard, and wildfire occurrence in coastal eastern Spain. Climatic change, 38(3), pp.345-357.\par

\bibitem[Pyne(2001)]{pyne2001}
Pyne, S.J., 2001. The fires this time, and next. Science, 294(5544), pp.1005-1006.\par

\bibitem[Radeloff \emph{et al}(2005)]{rade2005}
Radeloff, V.C., Hammer, R.B., Stewart, S.I., Fried, J.S., Holcomb, S.S. and McKeefry, J.F., 2005. The wildland–urban interface in the United States. Ecological applications, 15(3), pp.799-805.\par

\bibitem[Randerson \emph{et al}(2005)]{rand2005}
Randerson, J.T., Van der Werf, G.R., Collatz, G.J., Giglio, L., Still, C.J., Kasibhatla, P., Miller, J.B., White, J.W.C., DeFries, R.S. and Kasischke, E.S., 2005. Fire emissions from C3 and C4 vegetation and their influence on interannual variability of atmospheric CO2 and $\delta$13CO2. Global Biogeochemical Cycles, 19(2).\par

\bibitem[Reisen and Brown(2006)]{reis2006}
Reisen, F. and Brown, S.K., 2006. Implications for community health from exposure to bushfire air toxics. Environmental Chemistry, 3(4), pp.235-243.

\bibitem[Reus Dolz \emph{et al}(2003)]{reus2003}
Reus Dolz, M.L. and Irastorza, F., 2003. Estado del Conocimiento de causas sobre los incendios forestales en España. APAS \& IDEM Estudio sociologico sobre la percepcion de la pobalcion española hacia los incendios forestales.< www. idem21. com/descargas/pdfs/IncediosForestales. pdf.\par

\bibitem[Rodrigues and de la Riva(2014)]{rori2014}
Rodrigues, M. and de la Riva, J., 2014. An insight into machine-learning algorithms to model human-caused wildfire occurrence. Environmental Modelling \& Software, 57, pp.192-201.\par

\bibitem[Rodrigues \emph{et al}(2013)]{rodr2013}
Rodrigues, M., San Miguel, J., Oliveira, S., Moreira, F. and Camia, A., 2013. An insight into spatial-temporal trends of fire ignitions and burned areas in the European Mediterranean countries. Journal of Earth Science and Engineering, 3(7), p.497.\par

\bibitem[Rodrigues \emph{et al}(2014)]{rodr2014}
Rodrigues, M., de la Riva, J. and Fotheringham, S., 2014. Modeling the spatial variation of the explanatory factors of human-caused wildfires in Spain using geographically weighted logistic regression. Applied Geography, 48, pp.52-63.\par

\bibitem[Rodrigues \emph{et al}(2016)]{rodr2016}
Rodrigues, M., Jiménez, A. and de la Riva, J., 2016. Analysis of recent spatial–temporal evolution of human driving factors of wildfires in Spain. Natural Hazards, 84(3), pp.2049-2070.\par

\bibitem[Rue \emph{et al}(2009)]{rue2009}
Rue, H., Martino, S. and Chopin, N., 2009. Approximate Bayesian inference for latent Gaussian models by using integrated nested Laplace approximations. Journal of the royal statistical society: Series b (statistical methodology), 71(2), pp.319-392.\par

\bibitem[San-Miguel-Ayanz \emph{et al}(2010)]{sanm2010}
San-Miguel-Ayanz, J. and Camia, A., 2010. Forest Fires. In ‘Mapping the Impacts of Natural Hazards and Technological Accidents in Europe: an Overview of the Last Decade’. European Environment Agency Technical Report N, 13, pp.47-53.\par

\bibitem[San-Miguel-Ayanz \emph{et al}(2012)]{sanm2012}
San-Miguel-Ayanz, J., Rodrigues, M., de Oliveira, S.S., Pacheco, C.K., Moreira, F., Duguy, B. and Camia, A., 2012. Land cover change and fire regime in the European Mediterranean region. In Post-fire management and restoration of southern European forests (pp. 21-43). Springer, Dordrecht.\par

\bibitem[San-Miguel-Ayanz \emph{et al}(2013a)]{sanm2013a}
San-Miguel-Ayanz, J., Schulte, E., Schmuck, G. and Camia, A., 2013. The European Forest Fire Information System in the context of environmental policies of the European Union. Forest Policy and Economics, 29, pp.19-25.\par

\bibitem[San-Miguel-Ayanz \emph{et al}(2013b)]{sanm2013b}
San-Miguel-Ayanz, J., Moreno, J.M. and Camia, A., 2013. Analysis of large fires in European Mediterranean landscapes: lessons learned and perspectives. Forest Ecology and Management, 294, pp.11-22.\par

\bibitem[Schmid and Held(2004)]{schm2004}
Schmid, V. and Held, L., 2004. Bayesian extrapolation of space–time trends in cancer registry data. Biometrics, 60(4), pp.1034-1042.\par

\bibitem[Schrödle and Held(2011)]{schr2011}
Schrödle, B. and Held, L., 2011. Spatio‐temporal disease mapping using INLA. Environmetrics, 22(6), pp.725-734.\par

\bibitem[Sebastián-López \emph{et al}(2008)]{seba2008}
Sebastián-López, A., Salvador-Civil, R., Gonzalo-Jiménez, J. and SanMiguel-Ayanz, J., 2008. Integration of socio-economic and environmental variables for modelling long-term fire danger in Southern Europe. European Journal of Forest Research, 127(2), pp.149-163.\par

\bibitem[Team, R.C.(2013)]{team2013}
Team, R.C., 2013. R: A language and environment for statistical computing.\par

\bibitem[Turco \emph{et al}(2016)]{turc2016}
Turco, M., Bedia, J., Di Liberto, F., Fiorucci, P., von Hardenberg, J., Koutsias, N., Llasat, M.C., Xystrakis, F. and Provenzale, A., 2016. Decreasing fires in Mediterranean Europe. PLoS one, 11(3), p.e0150663.\par

\bibitem[Urbieta \emph{et al}(2019)]{urbi2019}
Urbieta, I.R., Franquesa, M., Viedma, O. and Moreno, J.M., 2019. Fire activity and burned forest lands decreased during the last three decades in Spain. Annals of Forest Science, 76(3), p.90.\par

\bibitem[Vázquez and Moreno(1998)]{vazq1998}
Vázquez, A. and Moreno, J.M., 1998. Patterns of lightning-, and people-caused fires in peninsular Spain. International Journal of Wildland Fire, 8(2), pp.103-115.\par

\bibitem[Vega-García \emph{et al}(1995)]{vega1995}
Vega-García, C., Woodard, P.M. and Lee, B.S., 1995. How GIS Can Help in Human Risk Rating and Daily Human-caused Forest Fire Occurrence Prediction. European Association of Remote Sensing Laboratories. Universidad de Alcalá.

\bibitem[Vilar Del Hoyo \emph{et al}(2009)]{vila2009}
Vilar Del Hoyo, L., Martin, P. and Camia, A., 2009. Analysis of human-caused wildfire occurrence and land use changes in France, Spain and Portugal. In Proceedings of the VII International EARSeL Workshop–Advances on Remote Sensing and GIS applications in Forest Fire Management. Potenza (Italy) (pp. 85-89).\par

\bibitem[Watanabe(2010)]{wata2010}
Watanabe, S., 2010. Asymptotic equivalence of Bayes cross validation and widely applicable information criterion in singular learning theory. Journal of Machine Learning Research, 11(Dec), pp.3571-3594.\par

\bibitem[Werth \emph{et al}(2016)]{wert2016}
Werth, P.A., Potter, B.E., Alexander, M.E., Clements, C.B., Cruz, M.G., Finney, M.A., Forthofer, J.M., Goodrick, S.L., Hoffman, C., Jolly, W.M. and McAllister, S.S., 2016. Synthesis of knowledge of extreme fire behavior: volume 2 for fire behavior specialists, researchers, and meteorologists. Gen. Tech. Rep. PNW-GTR-891. Portland, OR: US Department of Agriculture, Forest Service, Pacific Northwest Research Station. 258 p., 891.\par

\bibitem[Westerling \emph{et al}(2006)]{west2006}
Westerling, A.L., Hidalgo, H.G., Cayan, D.R. and Swetnam, T.W., 2006. Warming and earlier spring increase western US forest wildfire activity. science, 313(5789), pp.940-943.\par

\bibitem[Zumbrunnen \emph{et al}(2011)]{zumb2011}
Zumbrunnen, T., Pezzatti, G.B., Menéndez, P., Bugmann, H., Bürgi, M. and Conedera, M., 2011. Weather and human impacts on forest fires: 100 years of fire history in two climatic regions of Switzerland. Forest Ecology and Management, 261(12), pp.2188-2199.\par

\end{thebibliography}
\end{document}